\documentclass[conference,10pt,letterpaper]{IEEEtran}%
\usepackage{amsmath}
\usepackage{graphicx}
\usepackage{multirow}
\usepackage[none]{hyphenat}
\usepackage{float}
\usepackage{subfig}
\usepackage{dblfloatfix}
\usepackage{t1enc}
\usepackage{times}
\usepackage{siunitx}
\usepackage{cite}

\makeatletter

\def\@IMSauthorblockNAMEstyle{\normalfont\IMSauthorsize}
\def\@IMSauthorblockAFFILstyle{\normalfont\IMSaffilsize}
\def\@IMSauthorblockEMAILstyle{\normalfont\IMSaffilsize}
\def\IMSauthorblockNAME#1{%
\relax\@IMSauthorblockNAMEstyle%
#1%
}%
\def\IMSauthorblockAFFIL#1{%
\relax\@IMSauthorblockAFFILstyle%
\vskip\@IEEEauthorblockAtopspace
#1%
}%
\def\IMSauthorblockEMAIL#1{%
\relax\@IMSauthorblockEMAILstyle%
\vskip\@IEEEauthorblockAtopspace
#1%
}%
\newcommand{\IMSauthor}[1]{%
\ifIsBlindReviewVersion%
\author{\phantom{\parbox{\textwidth}{\center\relax#1}}}%
\else%
\author{\parbox{\textwidth}{\center\relax#1}}%
\fi%
}%
\newif\ifIsBlindReviewVersion
\def\IMSthispaperforblindreview{\IsBlindReviewVersiontrue}
\def\IMSthispaperforfinalpublication{\IsBlindReviewVersionfalse}
\def\@maketitle{\newpage
\bgroup\par\addvspace{0.5\baselineskip}\centering%
\ifCLASSOPTIONtechnote
   {\bfseries\large\@IEEEcompsoconly{\sffamily}\@title\par}\vskip 1.3em{\lineskip .5em\@IEEEcompsoconly{\sffamily}\@author
   \@IEEEspecialpapernotice\par{\@IEEEcompsoconly{\vskip 1.5em\relax
   \@IEEEtitleabstractindextextbox{\@IEEEtitleabstractindextext}\par
   \hfill\@IEEEcompsocdiamondline\hfill\hbox{}\par}}}\relax
\else
   \vskip0.2em{\IMStitlesize\ifCLASSOPTIONtransmag\bfseries\LARGE\fi\@IEEEcompsoconly{\sffamily}\@IEEEcompsocconfonly{\normalfont\normalsize\vskip 2\@IEEEnormalsizeunitybaselineskip
   \bfseries\Large}\@title\par}\vskip1.0em\par
   \ifCLASSOPTIONconference%
      {\@IEEEspecialpapernotice\mbox{}\vskip\@IEEEauthorblockconfadjspace%
       \mbox{}\hfill\begin{@IEEEauthorhalign}\@author\end{@IEEEauthorhalign}\hfill\mbox{}\par}\relax
   \else
      \ifCLASSOPTIONpeerreviewca
         {\@IEEEcompsoconly{\sffamily}\@IEEEspecialpapernotice\mbox{}\vskip\@IEEEauthorblockconfadjspace%
          \mbox{}\hfill\begin{@IEEEauthorhalign}\@author\end{@IEEEauthorhalign}\hfill\mbox{}\par
          {\@IEEEcompsoconly{\vskip 1.5em\relax
           \@IEEEtitleabstractindextextbox{\@IEEEtitleabstractindextext}\par\hfill
           \@IEEEcompsocdiamondline\hfill\hbox{}\par}}}\relax
      \else
         \ifCLASSOPTIONtransmag
           {\@IEEEspecialpapernotice\mbox{}\vskip\@IEEEauthorblockconfadjspace%
            \mbox{}\hfill\begin{@IEEEauthorhalign}\@author\end{@IEEEauthorhalign}\hfill\mbox{}\par
           {\vspace{0.5\baselineskip}\relax\@IEEEtitleabstractindextextbox{\@IEEEtitleabstractindextext}\vspace{-1\baselineskip}\par}}\relax
         \else
           {\lineskip.5em\@IEEEcompsoconly{\sffamily}\sublargesize\@author\@IEEEspecialpapernotice\par
           {\@IEEEcompsoconly{\vskip 1.5em\relax
            \@IEEEtitleabstractindextextbox{\@IEEEtitleabstractindextext}\par\hfill
            \@IEEEcompsocdiamondline\hfill\hbox{}\par}}}\relax
         \fi
      \fi
   \fi
\fi\par\addvspace{0.0\baselineskip}\egroup}

\def\IMStitlesize{\@setfontsize{\IMStitlesize}{18}{21pt}}
\def\IMSauthorsize{\@setfontsize{\IMSauthorsize}{12}{13pt}}
\def\IMSaffilsize{\@setfontsize{\IMSaffilsize}{12}{13pt}}
\def\IMScaptionsize{\@setfontsize{\IMScaptionsize}{8}{9pt}}
\def\IMSbibsize{\@setfontsize{\IMSbibsize}{8}{9pt}}

\def\@IEEEauthorblockNstyle{\IMSauthorsize\@IEEEcompsocnotconfonly{\sffamily}\@IEEEcompsocconfonly{\large}}
\def\@IEEEauthorblockAstyle{\IMSaffilsize\@IEEEcompsocnotconfonly{\sffamily}\@IEEEcompsocconfonly{\itshape}\@IEEEcompsocconfonly{\large}}
\def\@IEEEauthordefaulttextstyle{\IMSauthorsize\@IEEEcompsocnotconfonly{\sffamily}\sublargesize}

\def\thebibliography#1{\section*{\refname}%
    \addcontentsline{toc}{section}{\refname}%
    \IMSbibsize\@IEEEcompsocconfonly{\small}\vskip 0.3\baselineskip plus 0.1\baselineskip minus 0.1\baselineskip
    \list{\@biblabel{\@arabic\c@enumiv}}%
    {\settowidth\labelwidth{\@biblabel{#1}}%
    \leftmargin\labelwidth
    \advance\leftmargin\labelsep\relax
    \itemsep \IEEEbibitemsep\relax
    \usecounter{enumiv}%
    \let\p@enumiv\@empty
    \renewcommand\theenumiv{\@arabic\c@enumiv}}%
    \let\@IEEElatexbibitem\bibitem%
    \def\bibitem{\@IEEEbibitemprefix\@IEEElatexbibitem}%
\def\newblock{\hskip .11em plus .33em minus .07em}%
\ifCLASSOPTIONtechnote\sloppy\clubpenalty4000\widowpenalty4000\interlinepenalty100%
\else\sloppy\clubpenalty4000\widowpenalty4000\interlinepenalty500\fi%
    \sfcode`\.=1000\relax}

\long\def\@makecaption#1#2{%
\ifx\@captype\@IEEEtablestring%
\par\@IEEEtabletopskipstrut
\else
\@IEEEfigurecaptionsepspace
\fi
\setbox\@tempboxa\hbox{\normalfont\IMScaptionsize {#1.}\nobreakspace\nobreakspace #2}%
\ifdim \wd\@tempboxa >\hsize%
\setbox\@tempboxa\hbox{\normalfont\IMScaptionsize {#1.}\nobreakspace\nobreakspace}%
\parbox[t]{\hsize}{\normalfont\IMScaptionsize\noindent\unhbox\@tempboxa#2}%
\else
\ifCLASSOPTIONconference \hbox to\hsize{\normalfont\IMScaptionsize\hfil\box\@tempboxa\hfil}%
\else \hbox to\hsize{\normalfont\IMScaptionsize\box\@tempboxa\hfil}%
\fi\fi
\ifx\@captype\@IEEEtablestring%
\@IEEEtablecaptionsepspace
\else
\fi}

\newlength\tablecaptiontotableskip
\newlength\figuretocaptionskip
\def\@IEEEfigurecaptionsepspace{\vskip\figuretocaptionskip\relax}%
\def\@IEEEtablecaptionsepspace{\vskip\tablecaptiontotableskip\relax}%

\def\abstract{\normalfont%
\@IEEEabskeysecsize\bfseries\textit{\abstractname}\,\bfseries\textit{---}\,%
\@IEEEgobbleleadPARNLSP}%

\def\IEEEkeywords{\normalfont%
\@IEEEabskeysecsize\bfseries\textit{\IEEEkeywordsname}\,\bfseries\textit{---}\,%
\@IEEEgobbleleadPARNLSP}%
\def\endIEEEkeywords{\relax\vspace{0.67ex}%
\par\if@twocolumn\else\endquotation\fi%
\normalsize\normalfont}%

\DeclareRobustCommand*{\IMSauthorrefmark}[1]{\raisebox{0pt}[0pt][0pt]{\textsuperscript{\footnotesize{#1}}}}%
\def\@IEEEauthorblockNtopspace{0ex}
\def\@IEEEauthorblockAtopspace{1mm}
\def\IEEEkeywordsname{Keywords}
\def\subsubsection{\@startsection{subsubsection}{3}{\z@}{1.5ex plus 1.5ex minus 0.5ex}%
{0.7ex plus .5ex minus 0ex}{\normalfont\normalsize\itshape}}%
\def\@seccntformat#1{\csname the#1dis\endcsname\relax}
\def\thesubsectiondis{{\hbox to\parindent{\Alph{subsection}.}}}
\def\thesubsubsectiondis{{\hbox to \parindent{\arabic{subsubsection})}}}
\def\theparagraphdis{{\hbox to \parindent{\alph{paragraph})}}}
\IEEEilabelindentA \parindent
\IEEEilabelindent \IEEEilabelindentA
\IEEEelabelindent \parindent
\IEEEdlabelindent \parindent
\IEEElabelindent \parindent

\newlength\@IMSparindent
\newcommand\IMSdisplayacksection[1]{%
\ifIsBlindReviewVersion%
\noindent\phantom{\parbox[t]{\columnwidth}{\normalbaselines\setlength{\parindent}{\@IMSparindent}{#1}\strut}}
\else%
\noindent\parbox[t]{\columnwidth}{\normalbaselines\setlength{\parindent}{\@IMSparindent}{#1}\strut}%
\fi%
}%

\makeatother

\usepackage{tikz}
\usepackage{textcomp}
\usepackage{hyperref}
\usepackage{lipsum}

\newcommand\copyrighttext{%
  \footnotesize \textcopyright 2023 IEEE. Personal use of this material is permitted.
  Permission from IEEE must be obtained for all other uses, in any current or future
  media, including reprinting/republishing this material for advertising or promotional
  purposes, creating new collective works, for resale or redistribution to servers or
  lists, or reuse of any copyrighted component of this work in other works.
  }
\newcommand\copyrightnotice{%
\begin{tikzpicture}[remember picture,overlay]
\node[anchor=south,yshift=10pt] at (current page.south) {\fbox{\parbox{\dimexpr\textwidth-\fboxsep-\fboxrule\relax}{\copyrighttext}}};
\end{tikzpicture}%
}

\begin{document}
\raggedbottom
%
%
%
\title{Achieving Efficient and Realistic Full-Radar Simulations and Automatic Data Annotation by exploiting Ray Meta Data of a Radar Ray Tracing Simulator}
%
%
%
\IMSthispaperforblindreview
\IMSthispaperforfinalpublication
\IMSauthor{%
\IMSauthorblockNAME{
Christian Schüßler\IMSauthorrefmark{\#1*},
Marcel Hoffmann\IMSauthorrefmark{\#*},
Vanessa Wirth\IMSauthorrefmark{§*}, 
Björn Eskofier\IMSauthorrefmark{\$*}, 
Tim Weyrich\IMSauthorrefmark{$\dagger$*},
Marc Stamminger\IMSauthorrefmark{§*},
Martin Vossiek\IMSauthorrefmark{\#*}
}
\\%
\IMSauthorblockAFFIL{
\IMSauthorrefmark{\#}Institute of Microwaves and Photonics\\
\IMSauthorrefmark{§}Chair of Visual Computing\\
\IMSauthorrefmark{$\dagger$}Chair of Digital Reality\\
\IMSauthorrefmark{\$}Machine Learning and Data Analytics Lab\\
\IMSauthorrefmark{*}Friedrich-Alexander-Universität Erlangen-Nürnberg, Erlangen, Germany
}
\\%
\IMSauthorblockEMAIL{
\IMSauthorrefmark{1}christian.schuessler@fau.de
}
}
%

\maketitle
\copyrightnotice
%
%
%
\begin{abstract}
In this work a novel radar simulation concept is introduced that allows to simulate 
realistic radar data for Range, Doppler, and for arbitrary antenna positions in an efficient way.
Further, it makes it possible to automatically annotate the simulated radar signal by allowing to decompose 
it into different parts. This approach allows not only almost perfect annotations possible, but also allows 
the annotation of exotic effects, such as multi-path effects or to label signal parts originating from different parts
of an object.
This is possible by adapting the computation process of a Monte Carlo shooting and bouncing rays (SBR) simulator. 
By considering the hits of each simulated ray, various meta data can be stored such as hit position, 
mesh pointer, object IDs, and many more. This collected meta data can then be utilized to predict the change of 
path lengths introduced by object motion to obtain Doppler information or to apply specific ray filter rules
in order obtain radar signals that only fulfil specific conditions, such as multiple bounces or containing specific object IDs. 
Using this approach, perfect and otherwise almost impossible annotations schemes can be realized.
\end{abstract}
%
%

\section{Introduction}
Radar sensors have become one of the most important type of sensors when it comes to automated or autonomous driving applications.
Compared to lidar sensors, they are cheap, robust and can also operate under various weather 
conditions such as rain and fog~\cite{lit:RosiqueReviewAutomotiveSensors}.
Furthermore, radar sensors can directly measure the radial speed of objects by utilizing the Doppler effect.
Since radar sensors are especially unobtrusive and compact they can even be employed in domains such as human activity 
recognition~\cite{lit:Gurbuz2019ReviewRadarHumanActivity, lit:Li2019HumanActivityRecognition} or for medical tasks~\cite{lit:Liu2022Gaitmonitoring}.

In all of these domains, detection and classification for various cases plays a major role.
For example, entities in automotive scenarios such as pedestrians, cyclists, and cars have to be detected 
and classified reliably~\cite{lit:Sorowka2015PedestrianClassifcation, lit:Angelov2018AutomotiveClassification, lit:Prophet2018pedestrian}.
For activity recognition, various tasks such as hand gesture recognition~\cite{lit:Skaria2019HandGesture} or 
breathing and fall detection 
exist~\cite{lit:Adib2015smartBreathingAndHeartRate, lit:bhattacharya2020BreathingAndFall, lit:Amin2016radarElderlyFallDetection}.
Also the detection of ghost targets caused by multi-path reflections is important in all of these applications, 
since it can lead to false detections and classifications. Various 
machine learning approaches exist to alleviate this effect, 
see~\cite{lit:jin2021GhostTargetComparison, lit:Wang2021GhostTargetTransformer, lit:chamseddine2021ghost3D}.

In order to train potential classifiers, commonly the data has to be annotated beforehand.
This can either happen manually or self-supervised using reference sensors, 
such as lidar or camera sensors~\cite{lit:major2019DetectionWithLidar, lit:wang2021rodnetDetectionWithCamera}.
However, conducting real measurements and annotating data manually is expensive, time consuming, and often error-prone. 
Compared to natural images, radar data is very unfamiliar for the the human eye and expert knowledge is required 
for manual annotation.
Even with the help of reference sensors, the problems of generating enough corner cases remain. Further, neither lidar 
nor camera sensors share the same physical and data processing principles, which leads to an unreliable data base for self-supervised learning.

To overcome this problem, a digital twin of the real world and the sensor itself can be created~\cite{lit:dahmen2019digitalReality}.
This digital reality can serve as ground truth and allows for an improved annotation.
Even the car manufacturer Tesla, Inc. created a digital 
resemblance of San Francisco to train its autonomous driving algorithms\footnote[1]{https://www.teslaacessories.com/blogs/news/virtual-san-francisco-tesla\newline -tests-autopilot-with-simulation-from-large-game-engine-unreal}, which underlines the importance of this topic.
\begin{figure}
	\centerline{\includegraphics[width=0.4\textwidth]{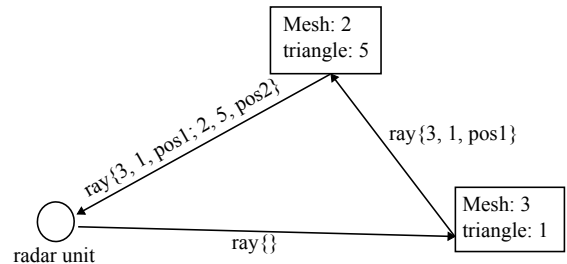}}
	\caption{A launched ray collects meta data in the simulation process, which can be used to generate Doppler simulations and efficiently annotate 
    radar images in the post processing step. In this illustration, the meta data consists of a a mesh id, a triangle index of the mesh, and the ray hit position.}
	\label{fig:RayConcept}
\end{figure}

For radar data, several simulation approaches exist, starting from point-scattering models~\cite{lit:buddendick2009radioPointScatterer}, 
generative machine learning approaches~\cite{lit:wheeler2017DeepStochasticModels} to very accurate ray tracing based physical models~\cite{lit:chipengo2020highAnsysSimulation, lit:thieling2020ScalableRadarSimulation, lit:hirsenkorn2017RayLaunching, lit:schussler2021RealisticRadarsimulation}.
Especially physical models can obtain very versatile and accurate simulations and 
are successfully used in classification tasks~\cite{lit:chipengo2021highSimulationClassification}. 
However, even in simulations it is hard to determine which part of the signal belongs to which specific object. 
This is because the reflectance at an object also causes sidelobes or creates multi-path targets besides the correct signal part. 
In the examplary case of a walking pedestrian obscured by a car, a ghost target can be created in the radar signal due to multi-path effects. 
For safety considerations it is essential to detect that
this signal actually originates from a human and, at the same time, that it is caused by multi-path effects.
Another simpler example is that even in simulations an object may be occluded by another one and therefore is not visible in the signal at all.
The simulator should therefore be able to detect this effect in order to avoid annotating this part in the radar image wrongly.

In this contribution, a concept is proposed that solves these issues by collecting ray meta data during the simulation process
with a radar ray tracing simulator. These meta data include pointers to the meshes and triangles that 
the rays hit on their way from transmitter to receiver. Moreover, 
the meta data can be augmented by even more attributes such as normal vectors or material information. This process is illustrated in Fig.~\ref{fig:RayConcept}.

With this meta data, it is possible to efficiently simulate Doppler characteristics as well as signals for large antenna arrays.
It is further possible to automatically generate (almost) perfect and precise labels for an annotation of the simulated radar signal.

\section{Concept}
\begin{figure}
	\centerline{\includegraphics[width=0.5\textwidth]{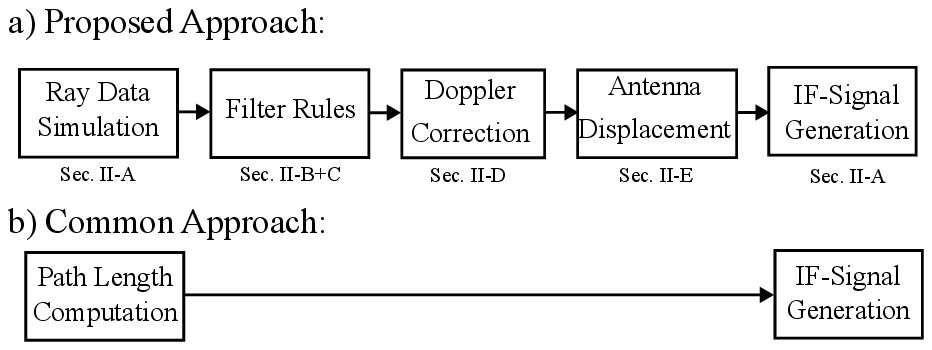}}
	\caption{The contribution of this work is shown in the upper part of the image (a). Compared to 
    conventional approaches (b) path lengths and ray data is only simulated once and are adjusted afterwards for
    different chirps and antenna positions. Below each box the respective section is referenced, which explains the processing step in more detail.}
	\label{fig:ChainConcept}
\end{figure}

In this section, the complete simulation workflow is explained from the radar signal model, to the utilization of ray meta data, up to the efficient simulation
for different antenna positions, as depicted in Fig.~\ref{fig:ChainConcept}.

\subsection{Radar Simulation And Signal Model}
The contribution of this paper is based on the simulator we presented in~\cite{lit:schussler2021RealisticRadarsimulation}.
It features an implementation of the shooting and bouncing rays (SBR) approach and uses a probabilistic material model to account for specular and diffuse reflection types. 
In order to account for realistic multi-path simulations and to achieve realistic material behavior, a large amount of rays is shot into the simulation
environment. This is computationally expensive but tolerable with modern hardware and a standard procedure for visual ray tracing approaches~\cite{lit:shirley2018ray}~\cite{lit:pharr2016physically}.

Conventionally, each ray with index $i$ sums up its path length $d_i$ from the transmitting (TX) antenna, 
to one or more objects and finally the receiving (RX) antenna. This process is also illustrated in Fig.~\ref{fig:RayConcept}. 
The delay $\tau_i$ for each ray is defined by 
\begin{equation}
    \label{eq:tau}
    \tau_i = \frac{d_i(j)}{c},
\end{equation}
with $c$ is being speed of light.
Due to the large number of rays, this information is sufficient to compute a good representation of the beat or intermediate frequency (IF) 
signal of a frequency modulated continuous wave (FMCW) radar signal as shown in the equation below 
\begin{equation}
    \label{eq:ifsignal}
    s_{\textrm{IF}}(t,j) = \sum^N_{i=0} A(\alpha, \beta) \exp(2\pi j(\mu t \tau_{i}(j) + f_c \tau_i(j))).
\end{equation}
With $A(\alpha, \beta)$ modelling the direction-depending antenna radiation pattern and $f_c$ representing the carrier frequency of the signal.
The frequency slope $\mu$ is defined by the ratio of the bandwidth $B$ and the chirp duration $T_c$, see equation below
\begin{equation}
    \label{eq:mu}
    \mu = \frac{B}{T_c}.
\end{equation}
A more detailed description of FMCW radar signal processing can for example be found in \cite{lit:li2021FMCWsignal}.

In order to simulate multiple Doppler shifts, so called \emph{chirps} have to be simulated. The time between two chirps is given
by $T_d$ and is often close to $T_c$ but not necessarily equal. 
Each chirp simulation is indicated by the index $j$ as used in~(\ref{eq:tau}) and~(\ref{eq:ifsignal}).
If an object is in motion during the radar measurement, the path length and therefore also $\tau$ will slightly change 
during each chirp. The time variable $t$ of a single chirp is 
commonly named \emph{fast time} and the time variable across all chirps with index $j$ is called \emph{slow time}. 
A complete measurement consisting of several chirps is called \emph{chirp sequence}.

Ordering the chirp sequence into a two dimensional array and computing a discrete Fourier transform results in a Range-Doppler spectrum
\begin{equation}
    S_{\textrm{IF}}(f_r,f_d) = \textrm{DFT}_{\textrm{2d}} (s_{\textrm{IF}}(t,j)).
\end{equation}\label{eq:range_doppler}
Zero padding and windowing is omitted in this short description. The resulting frequency axes represent the range frequency $f_r$ and the Doppler 
frequency $f_d$.
\subsection{Ray Meta Data}
In this subsection, the extension of the aforementioned simulation approach is described. 
The simulator will be adapted, so that it not only stores a ray's path length during the simulation, but also relevant additional information,
every time it hits a triangle.

\begin{figure}
	\centerline{\includegraphics[width=0.4\textwidth]{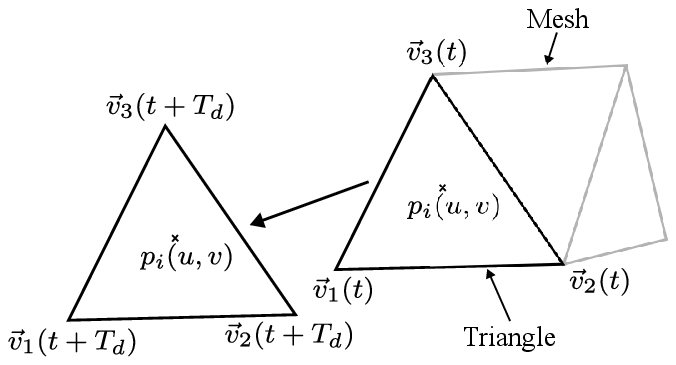}}
	\caption{This scheme shows an example of how a mesh with its assigned triangles moves through time. 
    After $T_d$ a new snapshot has to be taken and the hit position has to be updated. 
    Since the hit position and the hit triangle share the same transformation, 
    a single simulation run is sufficient to compute the barycentric coordinates $p_i(u,v)$ of the hit position.
    For all subsequent chirps, only the vertices have to be updated to compute the correct hit position.} 
	\label{fig:DopplerScheme}
\end{figure}

In general, the type and amount of meta data is arbitrary and only limited by the computing resources e.g.~video memory. 
However, in this work the meta data consists of a list of tuples. Each list entry represents a single ray hit and each tuple consists of the 
following elements: 
\begin{itemize}
    \item mesh ID $a_k$
    \item triangle index $b_k$
    \item barycentric triangle coordinates $u_k$ and $v_k$
\end{itemize}
Whereby, a  mesh represents an object consisting of 
multiple triangles and each triangle consists of three vertices, see Fig.~\ref{fig:DopplerScheme}.

The index \emph{k} determines the entry position in the list. Since the complete scene consists of several meshes, 
which again are made up of several triangles, identifiers for both have to be stored. To obtain the exact hit position 
after the ray tracing process, the barycentric coordinates of the hit position of the triangle have to be stored as well. 

It is also possible to store the Cartesian position of the hit position directly, but by using this approach, it 
simplifies the computation of Doppler information, as shown later in section~\ref{subsec:DopplerSim}.
Having a unique mesh ID is also required to decompose or to automatically label objects in the processed radar signal, 
as described in the next section.

\subsection{Radar Signal Decomposition}

Having stored all rays including their meta data, it is now possible to decompose the IF-Signal in the following way:
\begin{equation}
s_{\textrm{IF}}(t,j)  = \sum_{h=0}^{h=M} s^h_{\textrm{IF}}(t,j, \tau_h).
\end{equation}
Here, $h$ depicts a ray span or region that consists of several rays generating an IF-Signal part $s^h_{\textrm{IF}}$.
Each ray span may be generated by arbitrary filter rules, such as:
\begin{itemize}
    \item Any list entry that includes the mesh ID of a pedestrian
    \item Has more than one list entry to account for ghost-targets
    \item Has only one list entry and a specific mesh ID to account for line of sight detections of a specific object
    \item \ldots
\end{itemize}
With this technique, radar images including only specific properties can be created and labels can be automatically 
generated by applying very simple binary segmentation techniques in the processed radar image.

\subsection{Efficient Doppler Simulation}\label{subsec:DopplerSim}
Assuming that the motion across the slow time is small enough so that each ray would hit the same triangle at each chirp snapshot, 
one simulation run is sufficient as long as the initial ray hit positions are stored.
Since the position of each animated triangle is known at every specific point in time, 
the stored hit positions can be updated and new path lengths can be calculated.
This is much more efficient than running a complete simulation for each snapshot, because ray tracing requires a lot of expensive collision checks and most traced rays 
never hit any object or reach the receiver.
This idea has already been utilized in a similar way by~\cite{lit:azpilicueta2016intelligentDopplerSim} 
by considering the objects velocity and computing the Doppler shift directly. 
An extension of this was proposed in~\cite{lit:bilibashi2020DynamicRayTracing} for a image-based ray tracing simulation for multiple bounces.
However, this extension is implemented for an SBR approach and also supports non-linear motion in general. 
In fact, it still performs virtual snapshots but since the initial rays were already sampled, the remaining computation time turned out to be 
negligible.

This Doppler simulation process can be described as follows: after an initial ray sampling, the hit positions for each ray are known as barycentric coordinates $u$ and $v$. 
After the time $T_d$, the positions of each vertex of each animated object are updated. Since the hit positions should be moved by the same transformation as 
the vertices while still being placed on the triangle, the updated hit position can then the computed by

\begin{equation}
    \vec{p}_i = (1-u-v)\cdot \vec{v}_1 + u \cdot \vec{v}_2 + v \cdot \vec{v}_3,
\end{equation}

where the vectors $\vec{v}_1$,  $\vec{v}_2$, and $\vec{v}_3$ representing the three vertices of the stored triangle 
and $\vec{p}_i$ is the updated hit position in Cartesian coordinates after $T_d$. 
Updating each ray for the selected chirp, new path lengths and a new IF-Signal can be computed for each chirp.

\subsection{Simulating Different Antenna Positions}\label{subsec:BigArrays}
By assuming that the objects are sufficiently far away and the array aperture is small enough, 
path lengths for other antennas can be computed directly, as already done 
in a similar way in our initial work~\cite{lit:schussler2021RealisticRadarsimulation} for the displacement of TX antennas.
There, it was directly implemented in the ray tracing process, which is less efficient and consumes 
a significant higher amount of video memory.

Since the ray hit positions include the first and the last hit point, 
the path lengths caused by antenna displacement can easily be adjusted by the following formula

\begin{equation}
    l_i = l_i - |\vec{p}_i^{f} - \vec{x}_{tx}| + |\vec{p}_i^{f} - \vec{x}'_{tx}| -  |\vec{p}_i^{l} - \vec{x}_{rx}|+ |\vec{p}_i^{l} - \vec{x}'_{rx}|,
\end{equation}
with $\vec{p}_i^{f}$ being the first hit position and $\vec{p}_i^{l}$ being 
the last hit position for the simulated TX and RX antenna pair with positions $\vec{x}_{tx}$ and $\vec{x}_{rx}$.
The positions for the displaced antennas are denoted as  $\vec{x}'_{tx}$ and  $\vec{x}'_{rx}$, respectively.

\begin{figure}
	\centerline{\includegraphics[width=0.4\textwidth]{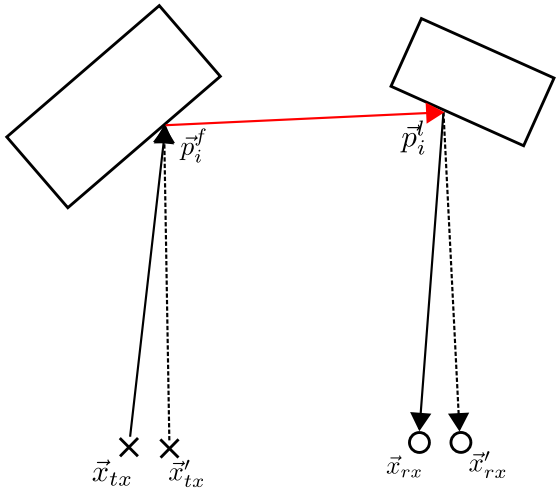}}
	\caption{The process of computing path lengths of multiple antennas without re-simulating the scene is depicted. Only the inner path length (red) remains 
    equal, path lengths directly connected to the antennas can be adjusted afterwards. This depiction should not be scaled as in typical real measurement scenarios the antennas are much closer and the objects 
    are placed much farer away from the antennas.} 
	\label{fig:MultipleAntennas}
\end{figure}

The idea is that, by subtracting the path lengths from the TX antenna to the first hit position and the path length from 
the RX antenna to the last hit position, only the path lengths, which originated through the environment remain. 
Assuming that the antenna array is small enough and objects are sufficiently far away, the reflection behavior of the objects 
does not change and the path lengths only have to be adjusted for the antenna displacement. This process is illustrated in Fig.~\ref{fig:MultipleAntennas}.
This approximation typically holds for automotive applications in the millimeter wave regime.

\section{Results}
\begin{figure*}
	\centerline{\includegraphics[width=0.8\textwidth]{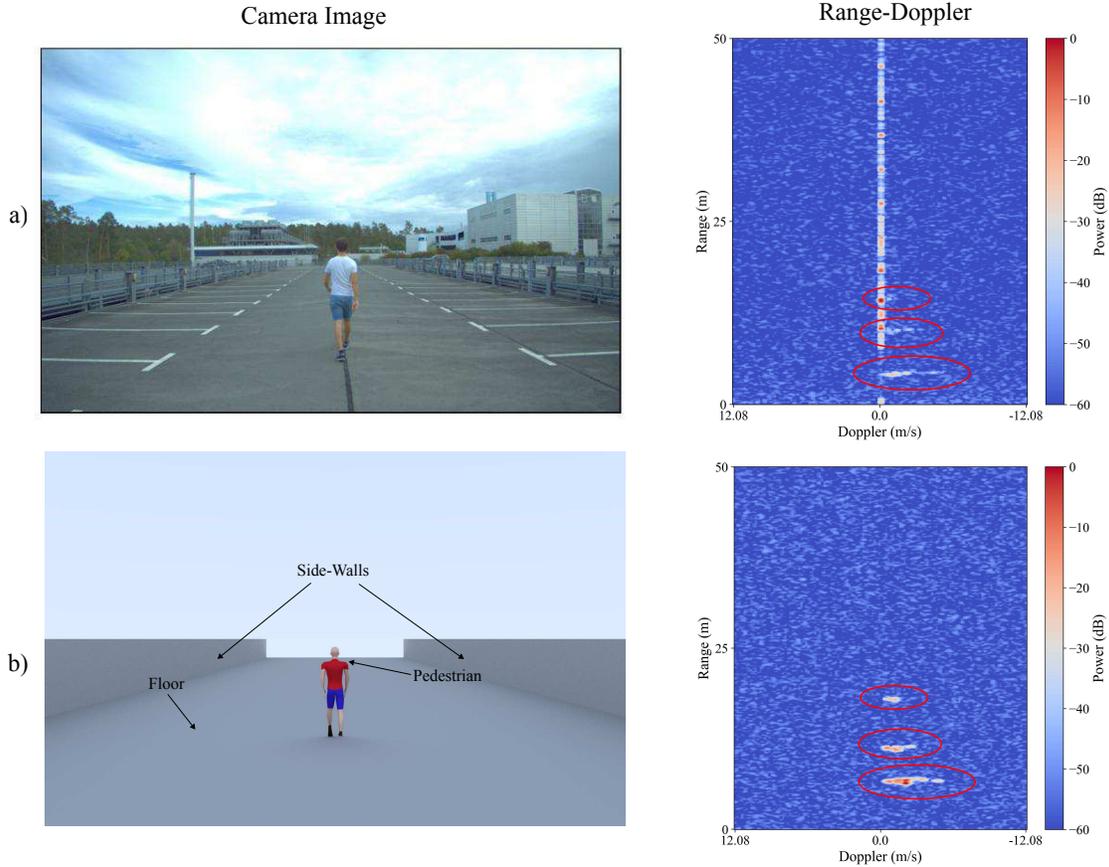}}
	\caption{The real measurement from a walking pedestrian is shown in the upper part of the figure (a) with its 
    simulated counterpart in the lower part of the figure~(b). As can be seen, both range-Doppler maps show the same 
    geometrical behavior and two ghost targets originating from multi-path effects are visible.
    Since the radar signal hits the walls in the simulation environment
    in a sharp angle, the signal is reflected away from the RX antennas making them invisible in the simulated radar image. 
    The walls in the real measurement are consisting of metal poles and bars, which can directly reflect the signal and are therefore visible.
    Since we focus mainly on multi-path effects, we did not take the effort to create such a detailed simulation environment.} 
	\label{fig:ResultsDoppler}
\end{figure*}

In the first experiment, we simulated a typical Range-Doppler map of a walking pedestrian and compared 
it with our measurement data. 
The measured and simulated range-Doppler image, as well as a camera image, can be seen in Fig.~\ref{fig:ResultsDoppler}. Clearly, a micro-Doppler signature
of the pedestrian can be seen, as highlighted by red ellipses. There also exist multi-path effects leading to 
ghost targets, which can also reproduced by the simulator. However, there are still some differences between simulation and measurement, which
might originate from an insufficient 3D description and from a too simple noise simulation.

\begin{figure*}
	\centerline{\includegraphics[width=0.9\textwidth]{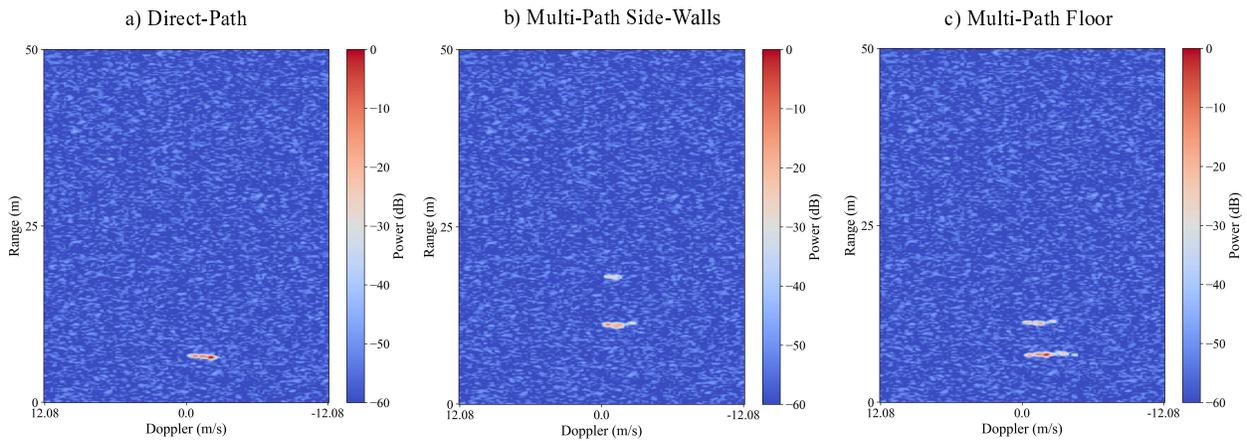}}
	\caption{Ray meta data was filtered in three different ways. In the left image (a) only direct rays without multiple bounces were considered during the signal generation.
    In the center image (b) rays with at least one bounce hitting the side-walls were used and in the right image (b) multi-path effects including the
    floor were considered.} 
	\label{fig:ResultMulti}
\end{figure*}

Since we can manipulate the ray meta data, 
it is possible to create a simulated range-Doppler image that only shows the ghost targets. As already mentioned, this is
especially helpful for machine learning algorithms to learn to differentiate between real and ghost targets for automotive applications.
This technique is demonstrated in Fig.~\ref{fig:ResultMulti} for different types of multi-path effects. 
As can be seen in image (c) the Doppler signal from the legs, 
assuming having the largest Doppler extent is mainly visible indirectly through reflections by the floor, while ghost-targets far away are generated by reflections from the side-walls.

\begin{figure}
	\centerline{\includegraphics[width=0.5\textwidth]{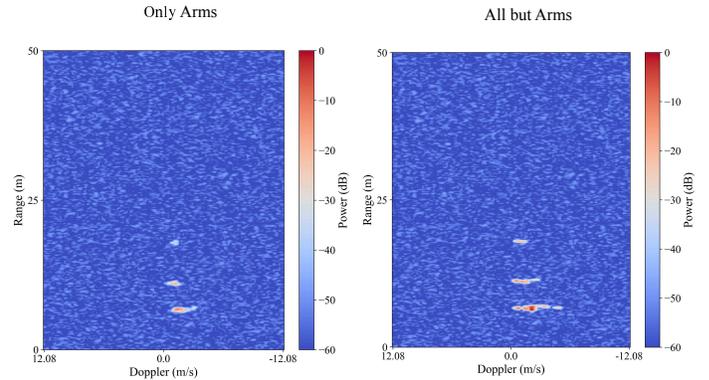}}
	\caption{In the left image, the complete scene was simulated as in Fig.~\ref{fig:ResultsDoppler}, but afterwards only rays that hit the arms of the pedestrian
    were used for the signal generation. In the right image, it is vice versa everything but the arms were kept during signal generation.} 
	\label{fig:ResultArms}
\end{figure}

As mentioned before, it is also possible to decompose the signal into different parts of an object. For example, in hand gesture recognition tasks only the signals 
reflected by hands or arms should be used for further processing. In Fig.~\ref{fig:ResultArms}, it is demonstrated how to separate the signal 
originated from the arms from the rest of the body. Here, the same pedestrian object and 3D environment was used as before.
As can be seen, the Doppler signal that only stems from the arms does overlap with the signal from the rest of the body but has 
significantly lower extend in Doppler direction, which is otherwise assumed to be caused by legs.

\section{Discussion and Future Work}
In this work, a radar simulation approach was presented that is based on a single TX and RX-antenna pair.
This is sufficient to generate simulated data for larger arrays and Doppler information. 
Consequently, a complete radar cube can be simulated. For objects, which are close 
to the antennas some assumptions may not hold since the reflection behavior of the surfaces is angle-dependent. 
In future work, also the normal vector of these surfaces shall be stored in the ray meta data, so that the received power for each antenna
may be adjusted accordingly.
Further, it was shown how the meta data of the simulated rays can be used to decompose the signal according to predefined filter rules. 
With this technique, simulation data can be automatically labeled in a way that would be mostly impossible by common simulation 
approaches and also extremely challenging for real measurement data.

Future work could potentially improve the simulation process by creating more diverse 3D worlds and compare the simulation with more complex measurement environments.
It should also be further evaluated to which extent simulation data can help to train machine learning algorithms.

\section*{Acknowledgment}


This work was supported in part by the Deutsche Forschungsgemeinschaft 
(DFG, German Research Foundation) under Grant SFB 1483–Project-ID 44241933.

\newcommand{\IMSacktext}{%
The authors would like to thank the Symeo team from
indie Semiconductor (Jannis Groh, Javier Martinez, and Mark
Christmann) for their support with the radar system used
for tests.
}

\IMSdisplayacksection{\IMSacktext}

\bibliographystyle{IEEEtran}
\bibliography{mybib}


\end{document}